# High Aptitude Motor-Imagery BCI Users Have Better Visuospatial Memory

Nikki Leeuwis
*Department of Cognitive Science and Artificial Intelligence*
*Tilburg University*
Tilburg, The Netherlands
N. Leeuwis@tilburguniversity.edu

Maryam Alimardani
*Department of Cognitive Science and Artificial Intelligence*
*Tilburg University*
Tilburg, The Netherlands
M.Alimardani@tilburguniversity.edu

*Abstract*— Brain-computer interfaces (BCI) decode the electrophysiological signals from the brain into an action that is carried out by a computer or robotic device. Motor-imagery BCIs (MI-BCI) rely on the user's imagination of bodily movements, however not all users can generate the brain activity needed to control MI-BCI. This difference in MI-BCI performance among novice users could be due to their cognitive abilities. In this study, the impact of spatial abilities and visuospatial memory on MI-BCI performance is investigated. Fifty-four novice users participated in a MI-BCI task and two cognitive tests. The impact of spatial abilities and visuospatial memory on BCI task error rate in three feedback sessions was measured. Our results showed that spatial abilities, as assessed by the Mental Rotation Test, were not related to MI-BCI performance, however visuospatial memory, assessed by the design organization test, was higher in high aptitude users. Our findings can contribute to optimization of MI-BCI training paradigms through participant screening and cognitive skill training.

I. INTRODUCTION

Brain-computer interfaces (BCI) convert brain signals into an action executed by a computer, enabling humans to interact with their surrounding environment via an external device and thus without using their muscles [1]. Brain activity is most commonly measured by electroencephalography (EEG) because compared to other imaging techniques, EEG is non-invasive, low-cost, and user-friendly [2].

In motor imagery BCIs (MI-BCI), subjects imagine moving a part of body, which generates changes in the brain activity of their motor cortex [3]. An algorithm is calibrated to recognize these changes and triggers an action based on the accuracy of the recognition [4]. Not all users can generate the brain activity needed to work with MI-BCI [5]. Previous studies indicate that between fifteen to thirty percent of MI-BCI users are entirely unable to generate the required brain activity patterns and are therefore incapable of controlling the system; this is called 'BCI illiteracy' [5] or 'BCI inefficiency' [4].

The inability to control the BCI device means that a user's performance does not meet the proficiency norm after the training. Lee et al. [2] showed that 55.6% of the first-time MI-BCI users scored below the 70% accuracy threshold during the first session. Several other studies confirmed this rate of MI-BCI illiteracy [6, 7, 8]. It is essential to investigate human factors in BCI performance because even the most advanced BCI classifier will not be able to discriminate brain activity patterns when the user cannot correctly modulate it.

Several studies have reported psychological, cognitive, and personal factors that influence MI-BCI performance (e.g., [6, 9, 10, 11, 12]). One of the most prevalent researches on the impacting factors of BCI performance is that of Jeunet et al. [9]. Amongst other variables, the researchers measured spatial ability (SA) with the Mental Rotation Test (MRT) [13], which was found to correlate to the BCI performance on a three-class Mental Imagery task. This effect was replicated by a follow-up study [6], which found a correlation between Mental Rotation and two-class MI-BCI peak performance but not with mean performance. Peak performance is calculated as performance during the time span where the accuracy of classification over all trials is highest, while mean performance represents the average accuracy during the complete feedback interval. Pacheco et al. [10] showed that Mental Rotation, among other SA tasks, were related to users' performance on a two-class MI-BCI which classified rest versus flexion of the arm. Teillet et al. [14] showed in a pilot study that training SA indeed could improve accuracy on a three-class mental imagery BCI. Pacheco et al. [10] also found other measures of SA that were related to BCI performance: spatial visualization ability as measured by the Block Design Test [15] was found to be corelated to BCI accuracy in a rest versus arm extension BCI task. In addition, Jeunet et al. [9] constructed linear models predicting BCI performance where an important predictor was the Corsi Block-Tapping Test [16], which measures memory consolidation of verbal and spatial sequences.

The current study extends the literature by exploring the relationship of spatial abilities and visuospatial memory to improve MI-BCI accuracy. The great advantage of the current study compared to the previous ones is the large number of participants, which provides statistical power and reliability of results. Jeunet et al. [9] included eighteen participants, Jeunet et al. [6] twenty and Pacheco et al. [10] contained only seven. Also, all studies that have shown an effect of spatial abilities on BCI performance adopted a BCI task that is inconsistent with previous literature (e.g., three-class mental imagery in Jeunet et al. [9] or rest versus arm flexion or extension in Pacheco et al. [10]). The current study used two-class motor imagery task, which is a common MI-BCI paradigm and is comparable to that of Jeunet et al. [6]. Spatial abilities were evaluated with the MRT [17] and the score was expected to impact BCI performance positively [9, 6, 10]. Visuospatial ability was assessed with the Design Organization Test [18] and was hypothesized to relate positively with MI-BCI performance [10]. Gender differences were found to impact both BCI-performance [19] and Mental Rotation scores [20] so the current study also evaluated the difference among genders to avoid any confounding effect.

## II. METHODS

### A. Participants

Fifty-seven participants were recruited for this study. Participants were students at Tilburg University and received either course credits or participated voluntarily. Only right-handed participants with (corrected to) normal vision who had never used a BCI before were eligible for the study. Two subjects were removed because they were not fit for the experiment. One participant exceeded the age range with 3 standard deviations (age = 39) and was therefore excluded. This left the experiment with 54 subjects ($M$ = 20.35, $SD$ = 2.41, 35 females, 19 males). The study was approved by the Research Ethics Committee of Tilburg School of Humanities and Digital Sciences (REDC #20201003). Prior to the experiment, participants read an information letter and singed an informed consent form.

### B. Cognitive Tests

Spatial ability and visuospatial memory were evaluated via two cognitive tests; Mental Rotation Test (MRT) and Design Organization Test (DOT).

MRT [17] assesses a subjects' ability to rotate objects within the mind. The MRT employed in the current study was a modified version of [21]. Stimuli were obtained from Peters et al. [20] and were presented in MATLAB 2019a (The MathWorks, Inc.) using the Psychtoolbox (Psychophysic Toolbox Version 3.0.16 [22]).

The stimuli consisted of two blocks of 40 trials, containing 20 normal and 20 mirrored stimuli. Before the start of the test, subjects had ten practice trials with feedback, however the actual trials in the task contained no feedback. The subject had to press the right arrow key to indicate that two stimuli were the same and the left arrow key to indicate that the stimuli were mirrored. Before each stimulus, a fixation cross was shown in the center of the screen for 0.25 seconds. The inter-trial interval was 0.25 seconds. If the trial was not answered within six seconds, the test automatically continued to the next trial.

The reference stimulus was a 30 degrees rotation around the x-axis relative to the original object because the original object had overlapping legs that caused the object to be unrecognizable in 2D space. The objects to compare with were rotated 80, 130, 190, 240, or 290 degrees around the x-axis relative to the original object (see Fig. 1). Eight reference stimuli were included. MRT scores were calculated as the percentage of correctly answered trials compared to the total number of trials.

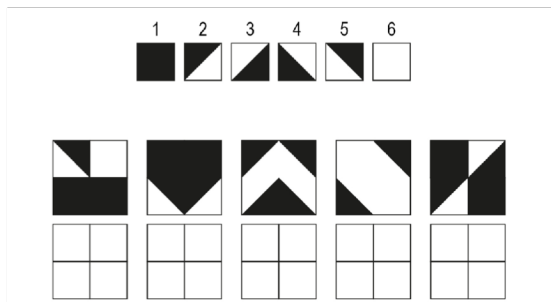

Fig. 2. An example of the Design Organization Test. The numbered squares at the top show the numerical code. The patterns in the squares below have to be reproduced by writing the number in the blank squares at the bottom.

DOT [18] quantifies visuospatial working memory and consists of a black-and-white square grid that display arrangements of full- or half-colored blocks. Participants were instructed to replicate as many designs as possible within one minute, using the numbered targets that function as puzzle pieces (see Fig. 2). The DOT consisted of one practice round and two test rounds. The scores are reflected by the sum of correct responses that the participant produced during the test rounds.

### C. BCI System and Motor Imagery Task

EEG signals were recorded from 16 electrodes distributed over the sensorimotor area according to the 10-20 system (F3,

Fig. 1. An example of the stimuli in the Mental Rotation Test. In every trial, a reference object is presented alongside a rotated object. The subject is required to determine whether the two objects are the same or different.

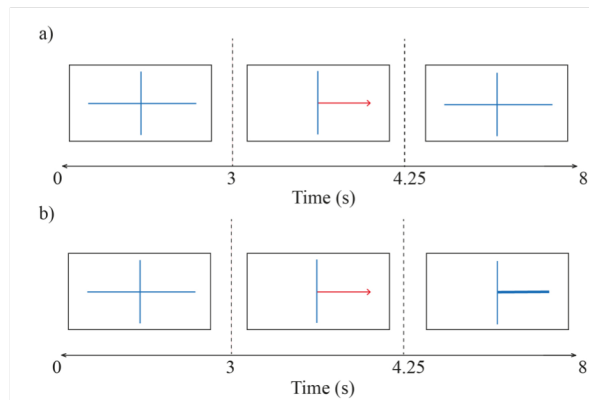

Fig. 3. The time course of BCI trials. (a) in the calibration run and (b) in the feedback runs. In all trials, participants saw a red arrow extending toward left or right side. They then had to imagine a movement of the corresponding hand. Only in feedback runs, they received feedback on their performance in form of a blue bar stretching toward left or right.

Fz, F4, FC1, FC5, FC2, FC6, C3, Cz, C4, CP1, CP5, CP2, CP6, T7, T8). A ground electrode was placed on the forehead (AFz) and a reference electrode was placed with a clip on the right earlobe. The EEG signals were amplified by a g.Nautilus amplifier (g.tec Medical Engineering, Austria). The sampling rate was 250 samples/second, and a 48-52 Hz Notch filter was applied to minimize the noise. In addition, a bandpass filter was employed from 0.5 to 30 Hz.

The BCI task consisted of four runs: one non-feedback run and three runs with feedback. Each run consisted of twenty left- and twenty right-hand trials of 8 seconds. The first run was the calibration run. The time course of events during a calibration trial can be found in Fig. 3a: each trial started with a fixation cross for 3 seconds, and thereafter a red arrow extending to left or right indicated the body side for which the participant had to imagine squeezing the hand. The arrow was presented for 1.25 seconds. After that, the fixation cross was visible for 3.75 seconds. The participant held the image of the corresponding movement in mind until a blank screen appeared, indicating the end of the trial. The time between trials was randomized between 0.5 and 2.5 seconds. After the calibration session, participants took a rest while the classifier parameters were calculated. Once the subject-specific classifier was set, participants were instructed to conduct the feedback runs. Each feedback run had 40 trials of 8 seconds. After the red arrow, a feedback bar was shown indicating the direction of the classifier's prediction (Fig. 3b). The length of the bar indicated the certainty.

Classification was done with the algorithms provided by g.tec. First, Common Spatial Patterns (CSP) was applied to extract spatial features of event-related (de-)synchronization during motor imagery and weigh the importance of each electrode for classification. Then, the obtained weight vectors were passed to a Linear Discriminant Analysis (LDA) classifier to both calibrate the system and discriminate between left- and right-hand movement imagination during feedback runs. Trials including artefacts were removed automatically.

*D. Experimental Procedure*

Before application for participation, subjects were informed about the content and procedure of the study by e-mail and completed the demographic questionnaire online. The demographic questionnaire attained background characteristics such as age, gender, and education and was administered using Qualtrics software (Qualtrics, I. (2013). Provo, UT, USA).

On the day of the experiment, the participant was seated in front of a desktop in a quiet room. Information on the procedure of the experiment was given to the participants before the participant signed the informed consent form. First, the DOT was taken using paper and pen and thereafter the MRT was administered on a laptop.

Once the cognitive tests were over, the experimenter placed the EEG cap and applied conductive gel. The impedance was kept below 50 kOhm. The experimenter explained that the participant had to avoid unnecessary movements during the recording and showed how brain activity would be contaminated due to noise caused by such events. Before the BCI task, the experimenter explained that the subject had to imagine squeezing the right or left hand without tension in their muscles. The subject tried a few trials before starting the calibration run to verify that they understood the motor imagery task. During feedback trials, subjects were requested to focus on imagining the movement and not get distracted by the feedback. One non-feedback calibration and three feedback runs were completed. The classifier was calibrated between every run based on data from the previous run. Once all runs were completed, the EEG cap was taken off and the subject was thanked for participating.

*E. Data Analysis*

The relationship of spatial ability and visuospatial memory with BCI performance was evaluated via two analyses; a correlation analysis on continuous data and a group comparison between high and low performers.

Continuous data included BCI error rates, which were obtained by g.BSanalyze software (g.tec Medical Engineering, Austria). The mean BCI error rate during the 8 second time window of all trails was computed for each of the three feedback-runs that the subjects completed. In order to specifically obtain the error rates during the MI task, a segment of the trials from 4.5 to 8 seconds was selected (see Fig. 3b). The segment duration was determined by making a consensus between Lee et al. [2] and Marchesotti et al. [11]. Finally, the average of BCI error rates in the segment of interest was obtained separately for each run and for all runs aggregated. This resulted in four dependent variables per subject: one BCI error rate for every run and one BCI error rate for all aggregated runs. BCI error rates and MRT scores were then checked for gender differences.

Correlation analyses were established between cognitive parameters and MI-BCI error rates. For normally distributed data, a Pearson correlation [23] was conducted and for non-normally distributed variables, Kendall coefficients [24] were obtained.

Besides the continuous BCI variable, performance was also evaluated as binary variable following Marchesotti et al. [11]. Subject were divided into two groups of high versus low aptitude users based on the median of the acquired BCI error rates. The difference of SA qualities in high and low aptitude BCI user groups in every run was compared. This was done using Welch's t-test [25] as the assumption of normality was met. When variables were not normally distributed, the groups were compared with a Mann-Whitney U-test [26]. Data analysis was done in R [27]. All checks for normality of the distributions were done with Shapiro-Wilk tests [28]. For all analyses, the significance level was maintained at 0.05.

III. RESULTS

For Mental Rotation, two subjects missed more than ten trials due to delayed answers; these were removed from analysis. There was no difference for gender on MRT score ($t(36.55) = .65, p = .52$).

DOT scores differed significantly between the first ($Mdn = 29.00, IQR = 4.00$) and the second ($M = 31.2, SD = 4.35$) round ($U = 930, p = .001$). The sum DOT score over both rounds was obtained ($M = 59.54, SD = 7.80$) and did not have any outliers. The distribution of BCI error rates can be found in Fig. 4. No effect of gender was observed in average BCI error rate ($U =$

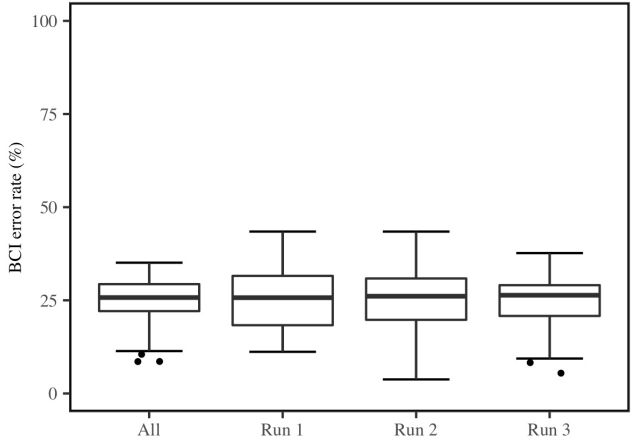

Fig. 4. Box plots of mean BCI error rates in every feedback run and all aggregated runs.

283, $p = 0.38$) between females ($Mdn = 25.7$, $N_F = 35$) and males ($Mdn = 26.2$, $N_M = 19$).

Correlation analysis revealed no significant correlations between BCI error rates and MRT scores or DOT scores (Fig. 5). The results can be found in Table I.

Fig. 6 illustrates MRT and DOT scores in high and low aptitude groups. Group comparison revealed no significant difference of Mental Rotation accuracy ($t(47.02) = .54$, $p = .59$) between high ($M = 69.33$) and low performers ($M = 67.40$). However, results of DOT showed that users who performed better in all BCI runs had significantly better visuospatial ability ($M = 61.93$) compared to low aptitude users ($M = 57.14$) ($t(52.00) = -2.35$, $p = .02$).

## IV. DISCUSSION

This study aimed to establish which spatial factors impact MI-BCI performance in novice users. Spatial abilities in the current study did not correlate with BCI performance. Similarly, visuospatial memory was not correlated with BCI performance, but it was shown that high MI-BCI performers had significantly greater visuospatial abilities.

Previous studies of Jeunet et al. [6, 9] and Pacheco et al. [10] found that spatial abilities as tested with the Mental Rotation Test (MRT) were correlated to MI-BCI performance. The current study did not show a correlation between BCI

TABLE I.

RESULTS OF CORRELATION ANALYSES BETWEEN BCI ERROR RATES AND COGNITIVE VARIABLES.

| BCI Run | MRT Score | DOT Score |
|---|---|---|
| Run 1 | $r = .08$, $p = .570$ | $r = -.12$, $p = .38$ |
| Run 2 | $r = .07$, $p = .64$ | $r = -.15$, $p = .30$ |
| Run 3 | $r = .04$, $p = .78$ | $r = -.13$, $p = .35$ |
| All Runs | $\tau = -.02$, $p = .80$ | $\tau = -.14$, $p = .14$ |

performance and MRT scores. Previous studies implemented the MRT by showing multiple objects at the same time [13], whereas the current research used the pairwise object presentation [17] because multiple object presentation tends to show gender differences [19]. Indeed, the scores on this implementation did not differ between the female and male participants of the current study. Another possible explanation for the difference in the obtained results could be that Mental Rotation was used in the study of Jeunet et al. [9] as part of the BCI task, which was executed together with a mental subtraction task and a left-hand motor imagery task (altogether 3 classes). Including MRT as a classification task may have driven the direction of the correlation found with MRT score; Christie et al. [29] showed that high MRT performers have stronger neural activations compared to low MRT performers. Friedrich et al. [30] compared the performance on various BCI tasks and found that left versus right motor imagery of the hand was most frequently demonstrating high performance, and this is also the most widely used BCI paradigm [31]. This motivated the implementation of a left versus right hand MI task in the current study.

Jeunet et al. [9] reported that the score of Corsi Block test was a predictor in linear models predicting BCI performance, suggesting that visuospatial memory is possibly related to BCI accuracy. The Corsi Block test [16] was replaced in this research because it is not fully understood what this test precisely evaluates [32, 33]. It is designed to measure memory consolidation of both verbal and spatial sequences but studies with patients with dissociation of visual and spatial memory show inconsistent results [33]. The current study incorporated the Design Organization Test (DOT), which is a reliable measure of visuospatial ability [34] and also strongly correlates with the commonly used and validated Block Design Test [15, 34], which was used by Pacheco et al. [10]. In the current study, the DOT was preferred over Block Design Test because the time of administration is much shorter [35]. Our results confirmed that high aptitude BCI users performed better on DOT than low aptitude users. Similar to our results, Pacheco et al. [10] showed that subjects' visuospatial abilities correlated positively with rest versus arm extension MI classification accuracy. It is worth noting that the results of both studies contradict that of Jeunet et al. [9] who used a Corsi Block Test and found that visuospatial memory span was a negative predictor of BCI performance. A more reliable test should establish the contribution of visuospatial factors in future research.

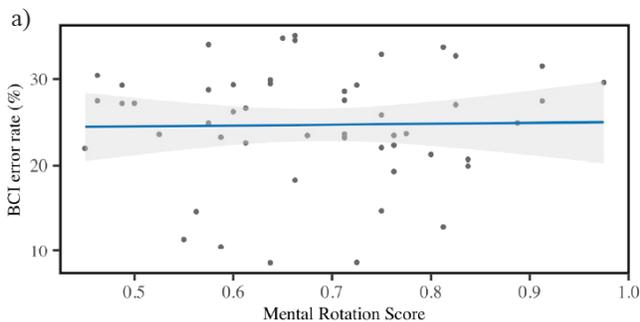
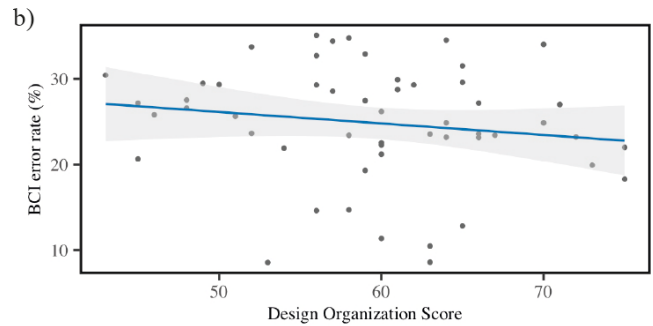

Fig. 5. Relation between the mean BCI error rates in all runs and cognitive variables; (a) Mental Rotation Score and (b) Design Organization Score

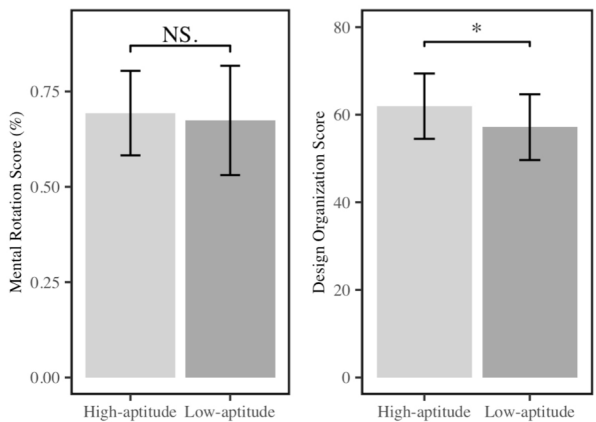

Fig. 6. Group comparisons of high and low aptitude BCI users on Mental Rotation and Design Organization Scores.

Most studies in the academic field quantify the relationship between BCI performance and cognitive measures using correlation analyses [9, 6, 10]. However, Machesotti et al. [11] divided the sample in high and low performers and the current study followed that example. Other studies took a similar approach but did not elaborate. For example, Jeunet et al. [6] picked the top ten and bottom ten users based on their performance in an earlier experiment for their comparison, which generated a wide gap between the BCI accuracies of the selected participants. In the current study, there were few extreme performers, but most were in the middle (i.e., their error rate was between 20 to 30%, see Fig. 4). Therefore, the approach of Machesotti et al. [11] was used alongside the common method of correlation analysis. The authors note that dividing the sample into two groups and comparing their characteristics is not a test of linear relationship, but we believe that establishing cognitive differences between high and low performers can provide a basis for further research on BCI (in-)efficiency.

The current study contributes to the literature by providing evidence on a larger sample; it evaluated spatial abilities in fifty-four participants while other studies have tested up to a maximum of twenty participants [6, 9, 10]. In addition, each of these studies reported an effect of spatial abilities on BCI performance, however they all employed a different kind of BCI classification task. Establishing the evidence on only one classification task strengthens the evidence and might affect the choice for future researchers.

Limitations of the current study include convenient sampling: this resulted in a skewed distribution of gender and age. In addition, the explanation given to the subjects about the task did not explicitly state whether the subject should imagine kinesthetic or visual movement, which is known to make a difference: kinesthetic imagery produces more distinguishable activation patterns in the primary motor cortex and supplementary areas because it is more intuitive compared to the visual imagery of movement [36]. Furthermore, the current study included only 120 feedback-trials, which might have reduced the validity. Typically, studies have 160 to 320 trials to produce reliable results [37]. When performance was evaluated per run, this number was reduced to forty trials.

Future research is recommended to draw a conclusion on the relationship between novice users' cognitive abilities and their MI-BCI performance. To make more robust predictions, it is needed to perform a large-scale study where subjects are trained for a more extended period. This would enable researchers to discriminate individual traits from states and observe when performance improves. Additionally, the BCI paradigm should be updated based on the latest research to make the study a reliable extension of the existing literature e.g., feedback with robot hands [38] or virtual reality [39]. As Teillet et al. [14] already showed in a pilot study, SA training might improve BCI performance. Improving the training conditions may reveal a more robust difference between (in-)efficient learners and thereby provide more valid evidence for the impacting variables on MI-BCI. Predicting whether the user is an efficient user or not would help decide the kind of training a user needs in order to work with motor imagery [40]. Adequate performance is essential for communication and control purposes [41]. Predicting BCI performance correctly saves researchers valuable time and enhances patient experience [12].

## V. CONCLUSION

The question remains whether cognitive abilities can predict MI-BCI performance. This study contributed to the literature by doubting the effect of Mental Rotation and visuospatial abilities. Our results showed that high aptitude users had a better visuospatial memory, however no effect was found for Mental Rotation and spatial abilities. The effect of visuospatial memory might be further explored and the finding that spatial abilities did not relate to BCI performance in the current study, might urge future research to explore the relationship even further. The large number of participants, the controlled environment in this experiment, the unique combination of measured variables and the use of the regular right versus left hand imagination instead of Mental Imagery tasks distinguish this study from previous research and make it a reliable reference for future research.

ACKNOWLEDGMENT

Authors would like to thank Alissa Paas for her assistance in collecting the data.